\documentclass[applemac]{aa}
\usepackage{natbib}
\usepackage{txfonts}
\usepackage{graphicx}
\usepackage{color}
\bibpunct{(}{)}{;}{a}{}{,}

\begin{document}

\title{Chemistry and dynamics of the prestellar core L1544}
\author{O. Sipil\"a,
	    P. Caselli,
	    E. Redaelli,
	    and S. Spezzano
}
\institute{Max-Planck-Institute for Extraterrestrial Physics (MPE), Giessenbachstr. 1, 85748 Garching, Germany \\
e-mail: \texttt{osipila@mpe.mpg.de}
}

\date{Received / Accepted}

\abstract{We aim to quantify the effect of chemistry on the infall velocity in the prestellar core L1544. Previous observational studies have found evidence for double-peaked line profiles for the rotational transitions of several molecules, which cannot be accounted for with the models presently available for the physical structure of the source, without ad hoc up-scaling of the infall velocity. We ran one-dimensional hydrodynamical simulations of the collapse of a core with L1544-like properties (in terms of mass and outer radius), using a state-of-the-art chemical model with a very large chemical network combined with an extensive description of molecular line cooling, determined via radiative transfer simulations, with the aim of determining whether these expansions of the simulation setup (as compared to previous models) can lead to a higher infall velocity. After running a series of simulations where the simulation was sequentially simplified, we found that the infall velocity is almost independent of the size of the chemical network or the approach to line cooling. We conclude that chemical evolution does not have a large impact on the infall velocity, and that the higher infall velocities that are implied by observations may be the result of the core being more dynamically evolved than what is now thought, or alternatively the average density in the simulated core is too low. However, chemistry does have a large influence on the lifetime of the core, which varies by about a factor of two across the simulations and grows longer when the chemical network is simplified. Therefore, although the model is subject to several sources of uncertainties, the present results clearly indicate that the use of a small chemical network leads to an incorrect estimate of the core lifetime, which is naturally a critical parameter for the development of chemical complexity in the precollapse phase.}


\titlerunning{L1544}
\maketitle

\section{Introduction}

Attempting to reproduce observed line emission (or absorption) is arguably the greatest possible challenge for an astrochemical model. The strengths and shapes of rotational lines as a function of frequency, or velocity, depend not only on the abundance of the molecule, but also on the physical structure of the source through the gas temperature and gas motions along the line of sight. In order to reproduce observations with an astrochemical model, one therefore also needs a detailed model for the physical structure including kinematic information.

L1544 is a prestellar core in the Taurus molecular cloud, located at a distance of 170\,pc \citep{Galli19}. It resides in a quiescent environment, has a high central density and a low central temperature, and shows clear signs of infall motions \citep{Tafalla98,Tafalla02,Crapsi07,Caselli12}, making it an excellent laboratory for investigating the properties of molecular gas in the final stages before the formation of a protostellar system. A one-dimensional model for the physical structure of L1544 has been presented by \citeauthor{Keto15}\,(\citeyear{Keto15}, hereafter denoted KC; see also \citealt{Keto10a}), based on a hydrodynamical simulation of the gravitational collapse of an object with L1544-like properties. The model yields the density, temperature, and infall velocity profiles as a function of distance from the center of the core, with the peak infall velocity reaching 0.14\,km\,s$^{-1}$. It has been used in several studies to represent L1544 in simulations of chemical abundance gradients and line emission \citep[e.g.,][]{Chacon-Tanarro19a,Redaelli19,Koumpia20,Caselli22}.

Single-dish observations toward L1544 have revealed double-peaked line profiles for several molecules, for rotational transitions that arise in intermediate-density ($10^4 < n < 10^6 \, \rm cm^{-3}$) gas: $\rm N_2H^+$ \citep{Bizzocchi13}, $\rm HCO^+$ \citep{Tafalla98}, $\rm HC^{17}O^+$ \citep{FerrerAsensio22}, and $\rm HC^{18}O^+$ \citep{Redaelli19}. The double peaks have been interpreted as being due to infall motions, and not because of multiple gas components on the line of sight. However, the infall velocity profile provided by the KC model has proven inadequate to explain the double-peaked lines, unless the velocity is scaled up. For example, \citet{Bizzocchi13} found that a velocity scaling of 1.75 provided the best fit to the observed $\rm N_2H^+$ lines.

The earlier works imply that the infall in L1544 is, in reality, proceeding at a faster rate than what has been previously thought based on the available physical model. The discrepancy could be explained by uncertainties as to the model. For example, the central density in the KC model is set under the assumption that the core is 140\,pc away \citep{Elias78}, which is $\sim$20\% less than the most recent distance measured with GAIA (and VLA) and reported in \citet{Galli19}. The targeted central density affects the choice of time step in the hydrodynamical model at which the physical properties are extracted from the simulation, which means that uncertainties as to this value directly affect the infall velocity as it is increasing rapidly during the collapse stage. The uncertainty as to the distance measurement also affects the mass estimate of the core.

However, uncertainties aside, there is also reason to believe that the effect of chemistry on the collapse dynamics, and hence on the infall velocity, may have been previously underestimated. We have shown in \citeauthor{Sipila18}\,(\citeyear{Sipila18}; hereafter S18) that considering a very large chemical network and a large set of molecular coolants can reduce the collapse timescale by a factor of two or more, when compared to simulations adopting a simple description of chemistry (as was done by KC). In the present paper, we aim to investigate the effect of chemistry on the infall velocity by running hydrodynamical simulations comparable to those carried out by KC. Line cooling powers are determined time-dependently based on the evolving molecular abundances, using radiative transfer simulations. We produce simulated line profiles to check for infall signatures in the rotational lines of several molecules, comparing these also with the observed counterparts. By running multiple simulations with different approaches to chemistry and line cooling, we can quantify the magnitude of the effect that the chemistry has on the dynamics of the collapse, and specifically to investigate the way in which changes in the chemical network or in the approach to determine the line cooling powers affect the infall velocity -- of particular interest is to determine if extending the chemical network can naturally lead to a higher infall velocity or if differences in simulation results can be attributed to radiative transfer effects.

The paper is organized as follows. In Sect.\,\ref{s:model}, we present the details of our simulation setup, including the hydrodynamical model as well as the chemical and radiative transfer models, and the initial conditions. The results of our simulations are presented in Sect.\,\ref{s:results}, and in Sect.\,\ref{s:discussion} we discuss the implications of our results along with an analysis of the uncertainties in the model setup. We give our conclusions in Sect.\,\ref{s:conclusions}. Appendix~\ref{a:optimization} describes an optimization scheme that enables a significant speed-up of the simulations presented in this paper.

\section{Model}\label{s:model}

In this section we present the details of the modeling setup, including details of the hydrodynamical, chemical, and radiative transfer simulations, as well as of the initial conditions. We also discuss some optimizations that we have made to the hydrodynamical model that enable a significant speed-up of the simulations.

\subsection{Model framework}\label{ss:modelFramework}

We employ our collapse model introduced and discussed in S18 (see also \citealt{Sipila19a}), here entitled HDCRT (short for HydroDynamics with Chemistry and Radiative Transfer). A detailed account of the model including the relevant equations is presented in S18; here we give only a brief summary. The model consists of a combination of hydrodynamical, chemical, and radiative transfer codes. The hydrodynamical code solves the hydrodynamics equations in a one-dimensional (1D) Lagrangian spherically symmetric framework, assuming that the cloud is supported purely by thermal pressure. The chemical code resolves the chemical evolution time-dependently based on the changing physical conditions, and the line cooling efficiencies of a variety of cooling molecules are computed time-dependently as a function of the chemical evolution via radiative transfer simulations. The combination of the different codes allows to track the physical and chemical evolution in the collapsing cloud completely self-consistently. In fact it was found in S18 that considering the effect of time-dependent gas cooling has a large effect on the collapse timescale because the gas is mostly cooled by molecular line radiation\footnote{Gas-dust collisional coupling dominates the cooling at high volume densities, while in the outer cloud which is exposed to the ISRF, external heating overpowers line cooling in determining the gas temperature.}.

The version of HDCRT used here is essentially the same as the one used in our two previous papers, although several refinements and optimizations have been made to the workflow which drastically reduce the required computational time (see Appendix~\ref{a:optimization}). We use the gas-phase and grain-surface chemical networks described in \citeauthor{Sipila19b}\,(\citeyear{Sipila19b}; full scrambling case). The initial chemical abundances are displayed in Table~\ref{tab:initialabundances}. We assume here monodisperse grains with radius $a_{\rm g} = 0.1 \, \mu \rm m$ with a grain material density of $\rho = 2.5 \rm \, g \, cm^{-3}$. The dust opacity is taken from \citet{Ossenkopf94}, corresponding to thin ice mantles, while the external radiation field is adopted from \citet{Black94}. We employ in our simulations a $G_0$ factor of 0.5; $G_0$ denotes the scaling of the strength of the external radiation field. This choice is motivated in Sect.\,\ref{ss:uncertainties} The visual extinction at the edge of the model cloud is set to 1\,mag. In addition to the cooling molecules listed in S18, we additionally include here, for the sake of completeness, cooling by $\rm H_2$ (\citealt{Wan18}; see also \citealt{Flower21}). Molecular hydrogen is expected to contribute to the total cooling power at low volume densities and high temperatures. In the present case it has however hardly any influence on the results, because the cooling powers of $\rm H_2$ and $\rm C^+$ become comparable only at temperatures above 100\,K, and our simulations predict clearly lower temperatures in the outer core.

\begin{table}
        \centering
        \caption{Initial abundances (with respect to $n_{\rm H} \approx 2\,n({\rm H_2})$) used in the chemical modeling.}
        \begin{tabular}{l|l}
                \hline
                \hline
                Species & Abundance\\
                \hline
                $\rm H_2$ & $5.00\times10^{-1}\,^{(a)}$\\
                $\rm He$ & $9.00\times10^{-2}$\\
                $\rm C^+$ & $1.20\times10^{-4}$\\
                $\rm N$ & $7.60\times10^{-5}$\\
                $\rm O$ & $2.56\times10^{-4}$\\
                $\rm S^+$ & $8.00\times10^{-8}$\\
                $\rm Si^+$ & $8.00\times10^{-9}$\\
                $\rm Na^+$ & $2.00\times10^{-9}$\\
                $\rm Mg^+$ & $7.00\times10^{-9}$\\
                $\rm Fe^+$ & $3.00\times10^{-9}$\\
                $\rm P^+$ & $2.00\times10^{-10}$\\
                $\rm Cl^+$ & $1.00\times10^{-9}$\\
                $\rm HD$ & $1.60\times10^{-5}$\\
                \hline
        \end{tabular}
        \label{tab:initialabundances}
        \tablefoot{$^{(a)}$ The initial $\rm H_2$ ortho/para ratio is $1 \times 10^{-3}$.}
\end{table}

Our fiducial simulation assumes initial conditions that are close, but not identical, to those used by KC. We adopt for the initial core structure an isothermal, unstable Bonnor-Ebert sphere \citep{Bonnor56,Ebert55}, with central density $n({\rm H_2}) = 2 \times 10^4 \, \rm cm^{-3}$, outer radius $1.1 \times 10^5 \, \rm au$, and mass $10\,M_{\sun}$. The outer radius and the total mass correspond to a temperature and a nondimensional radius of 10\,K and 23, respectively. Therefore our initial core is smaller than the KC one, and the average density is in our model about a factor of~6 higher. It is possible to obtain a larger initial core by decreasing the temperature and increasing the nondimensional radius while still constraining the mass to $10\,M_{\sun}$, but reaching an outer radius of $2 \times 10^5 \, \rm au$ implies a temperature of 6.5\,K throughout the initial core (because the core mass $M \propto T^{3/2}$; \citealt{Bonnor56}), which is not a very reasonable initial condition given that such low temperatures are expected only in very high-density gas. We have however run one simulation using such a larger initial core configuration (see Sect.\,\ref{ss:simulationCases}).

The HDCRT simulation is designed to run until one of the following three conditions is fulfilled: 1) the infall velocity exceeds twice the sound speed at any point in the cloud; 2) the time step (see below) becomes shorter than $10^2\,\rm yr$; 3) the simulation time exceeds a maximum of $10^7\,\rm yr$. Condition~1 follows from S18 where we imposed a similar condition simply to prevent the hydrodynamical solution from proceeding too far. Here, conditions 1~and~2 together ensure that the simulation is stopped before the central density climbs much above $n({\rm H_2}) = 10^7 \, \rm cm^{-3}$, which leads to numerical issues in the central core with the present version of the code, related to the determination of the line cooling powers at very small scales. In practice, most of our simulations terminate due to condition~1~or~2 -- whichever of these occurs first depends on the simulation parameters, for example the mass of the core.

\subsection{Simulation cases}\label{ss:simulationCases}

We run our hydrodynamical simulations until one of the termination conditions described in Sect.\,\ref{ss:modelFramework} is met. One fundamental aspect of the simulations is the method with which line cooling is implemented. For their model of L1544, KC used the parametrized cooling functions from \citet{Goldsmith01} and \citet{Tielens05}, whereas we calculate the line cooling power in a self-consistent and time-dependent manner using radiative transfer simulations (for this we employ the LOC program; \citealt{Juvela20}). The description of chemistry may affect the line cooling powers as a function of time. Also, expansion motions do not appear in the KC model, whereas we obtain an expansion of the outer core in our fiducial simulations as we discuss below.

To be able to evaluate how the abovementioned issues affect the simulation results, we have run a total of five hydrodynamical simulations. Our fiducial simulation~HD0 corresponds to the initial parameters described in Sect.\,\ref{ss:modelFramework}, and includes a full treatment of line cooling based on radiative transfer without restrictions to the dynamics of the collapse. A second simulation (HD1) is otherwise identical to HD0, but the core is not allowed to expand. Here we enforce the condition $\min(\varv,0)$ for the gas velocity at each point in the cloud. Two further simulations switch, respectively, to a simpler chemical scheme (HD2) and to parametrized line cooling functions (HD3; we discuss the cooling parametrization more in Sect.\,\ref{ss:alternativeModels}). Both of these simulations utilize the B2alt chemical network from S18 which is similar to the network used by KC, although it includes CO and $\rm H_2O$ formation on grains which KC did not consider. The simulation cases are summarized in Table~\ref{tab:simulationCases}. By sequentially simplifying the model we can determine if, and how, the various properties of the simulation affect the results.

To explore the effect of the average density of the initial core on the simulation results, we have also run one hydrodynamical simulation (HD4) which is otherwise identical to HD3 but adopts the larger initial core that extends to $2 \times 10^5 \, \rm au$ in radius. This is achieved by setting the temperature of the initial core to 6.5\,K and the nondimensional radius to 50.

In past works where we have attempted to reconcile the results of chemical simulations with observations, we have employed a so-called static model where the physical structure of the core remains unchanged while the chemistry is evolving. As discussed in S18, the results of dynamical and static models may differ greatly as a function of location in the core. We have therefore run two static simulations for the present paper as well. In both cases, we run a time-dependent chemical simulation (including self-shielding) up to the time of best match from simulation~HD0 ($9.77 \times 10^5 \, \rm yr$; see Sect.\,\ref{ss:fiducialModel}) using the same initial conditions as in HD0. The only difference between the static simulations is that one adopts the physical structure from HD0, while the other adopts the corresponding profiles from KC. In these two simulations a constant time step of $3.0 \times 10^3 \, \rm yr$ is used, that is, 334 time steps over a period of $10^6 \, \rm yr$. This step is somewhat longer than the initial time step in the hydrodynamical simulations ($2.3 \times 10^3 \, \rm yr$), but still the total number of steps taken over the course of core evolution is a factor of several higher than what we normally consider in our static simulations, such as those presented in \citet{Sipila19a}, over a similar period of simulation time.

To facilitate the comparison of our results to observations, we also carry out radiative transfer simulations to produce simulated line emission based on the results of the dynamical and static models (using LOC). We adopt a source distance of 170\,pc and set the per channel velocity resolution to match those in the observations of the various lines (see Sect.\,\ref{ss:lineSimulations}). All line simulations adopt a constant 0.05\,km\,s$^{-1}$ level of turbulence across the core.

\begin{table}
        \renewcommand{\arraystretch}{1.3}
        \centering
        \caption{Summary of simulation cases discussed in the text.}
        \begin{tabular}{c | c }
                \hline
                \hline
                Simulation & Description \\
                \hline
                HD0 & Fiducial hydrodynamical simulation \\ 
                \hline
                HD1 & As HD0, but expansion motions are not \\
                		&	allowed \\
                \hline
                HD2 & As HD1, but adopting the B2alt chemical\\
		        &      network presented in S18 \\ 
                \hline
                HD3 & As HD2, but employing a parametrized line \\
                		&	cooling scheme \\
                \hline
                HD4 & As HD3, but using a larger initial core that \\
                      & extends to $2 \times 10^5 \, \rm au$ \\
                \hline
                ST0 & Static physical model adopting the parameters \\
                	     & and physical structure from simulation~HD0  \\ 
                \hline
                STKC & As ST0, but adopting the KC physical structure \\
                \hline
        \end{tabular}
        \label{tab:simulationCases}
\end{table}

\section{Results}\label{s:results}

We present here the results of the simulations introduced above. Even though we cannot match the simulation setup of KC precisely, we still aim to compare our results to theirs as closely as possible. To this end, we focus in all of our hydrodynamical simulations on a time at which the central density of the core matches best the central density in the KC model, that is, $n({\rm H_2}) = 8.3 \times 10^6 \, \rm cm^{-3}$ -- this time is the ``time of best match'' ($t_{\rm bm}$) mentioned in Sect.\,\ref{ss:simulationCases} above, and varies from simulation to simulation. The actual central density in L1544 and hence the best-fit time in the simulations is of course subject to uncertainties, which we discuss in Sect.\,\ref{ss:uncertainties}.

\subsection{Hydrodynamical simulations HD0 and HD1}\label{ss:fiducialModel}

\begin{figure*}
\centering
        \includegraphics[width=2.0\columnwidth]{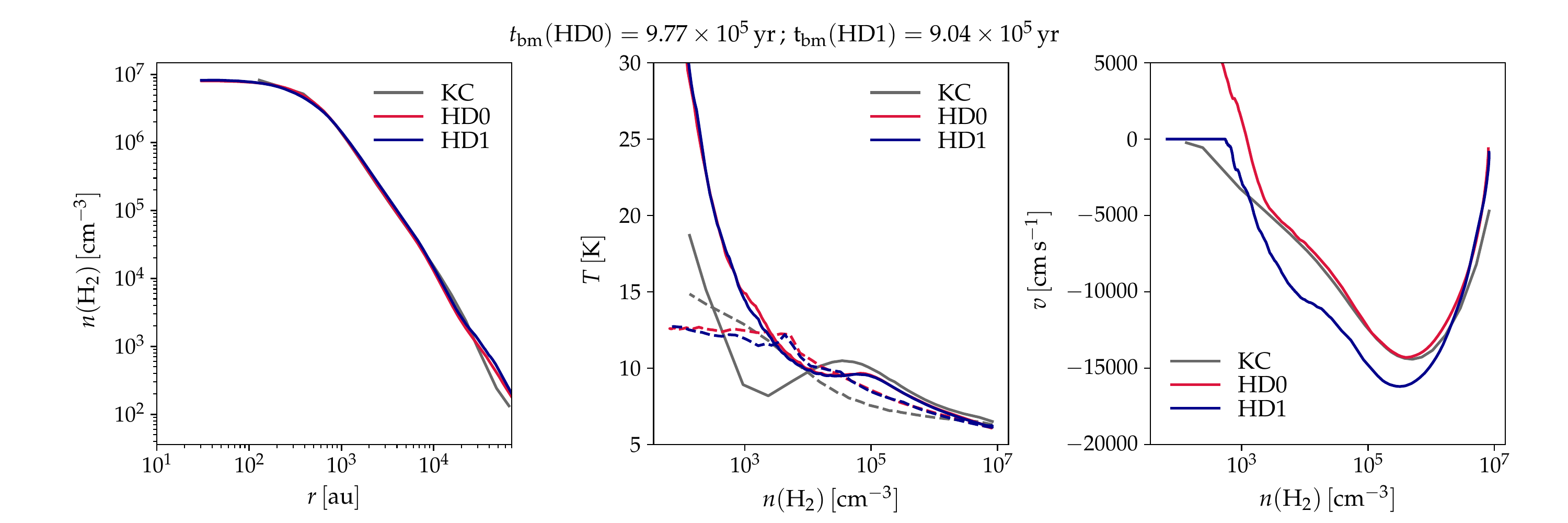}
    \caption{Comparison of the physical structure at $t_{\rm bm}$ in simulations HD0 (red) and HD1 (blue). The left panel shows the volume density as a function of radius. In the middle panel, the solid and dashed lines represent the gas and dust temperature, respectively, shown as a function of volume density. The right panel shows the infall velocity as a function of volume density. The gray curves show the corresponding data from the KC model. The values of $t_{\rm bm}$ in simulations HD0 and HD1 are shown on top.}
    \label{fig:physicalStructureFiducial}
\end{figure*}

The fiducial simulation HD0 terminated at a simulation time of $9.82 \times 10^5$\,yr. At this time, the central density of the core was $2.6 \times 10^7 \, \rm cm^{-3}$. The best match to the KC central density is obtained some tens of kyr earlier, at a time $t_{\rm bm}({\rm HD0}) = 9.77 \times 10^5$\,yr. In simulation~HD1, the corresponding values are $9.09 \times 10^5$\,yr and $t_{\rm bm}({\rm HD1}) = 9.04 \times 10^5$\,yr. Figure~\ref{fig:physicalStructureFiducial} shows the density, temperature, and infall velocity profiles at $t_{\rm bm}({\rm HD0})$, along with the corresponding results from simulation~HD1 at $t_{\rm bm}({\rm HD1})$. We also show the KC infall velocity profile for comparison. There are interesting similarities, and differences, between the results.

Firstly, the density profiles predicted by the two HDCRT simulations and the KC model are very similar. One can note that the volume density in the outer core is somewhat higher in the present simulations as compared to the KC model, which is a consequence of the average mass of the core being higher; most of the core mass is in the outer layers.

Secondly, the dust and gas temperature profiles are nearly identical between the two HDCRT simulations. However, HDCRT predicts a consistently higher dust temperature compared to KC, except near the edge of the core. The dust temperature is affected by the density profile, by the radiation field incident on the core, and of course by the dust opacity model. HDCRT and the KC model adopt the same spectrum for the interstellar radiation field \citep{Black94} and source data for the dust opacity from \citeauthor{Ossenkopf94}\,(\citeyear{Ossenkopf94}; see \citealt{Keto05}, \citealt{Zucconi01}). Given that the density profiles are essentially the same between the models, the spatial difference in dust temperature is thus attributed to differences in the assumptions of extinction external to the core, and to different opacity parameters\footnote{KC multiplied the \citet{Ossenkopf94} dust opacities artificially by a factor of 4.}. The gas temperature in the inner core is determined by collisional coupling with the dust. The HDCRT and KC models predict very similar gas temperatures up to $\sim$6000\,au away from the core center. When the gas density drops to below approximately $10^5\,\rm cm^{-3}$, the dust and gas become collisionally decoupled and the gas temperature decreases owing to molecular line cooling. Near the core edge the temperature rises again due to external heating -- both models predict a local minimum in the gas temperature, which traces the zone where the dust-gas collisional coupling is weak and the external radiation is attenuated, that is, where the effect of molecular line cooling is the greatest. The gas temperature starts tending toward the minimum at a smaller radius in the present model despite the very similar density profiles in the two HDCRT simulations and in the KC model. The gas temperature also turns to a strong increase at small radii in simulations HD0 \& HD1, implying again differences in the treatment of dust (via photoelectric heating) and extinction.

Regarding the collisional coupling, it is also evident in Fig.\,\ref{fig:physicalStructureFiducial} that the coupling between the gas and the dust is much weaker in the KC model as compared to HDCRT. In the former model, the two temperatures are approximately equal only in the innermost $\sim$700\,au of the core, while in the latter the region of approximate equality extends to $\sim$2000\,au. The origin of the discrepancy is unknown -- both models use the collisional coupling scheme of \citet{Goldsmith01} and the same dust-to-gas mass ratio (100).

Thirdly, the infall velocity is, at $t_{\rm bm}$, quite different between the two HDCRT simulations. In particular, the infall velocity is higher almost throughout the core in the HD1 simulation, as compared to HD0. Though the peak of the velocity profile is co-located in both simulations, the peak is broader in HD1. We recall that the only difference in the HD0 and HD1 simulation setups is that in the latter the gas is not allowed to expand. The expansion motion evident in the outer core in simulation~HD0 is a consequence of the strong warming up of the gas due to photoelectric heating, which causes the gas to expand. The absence of expansion means that the fraction of the gas that is infalling is higher in simulation~HD1, leading to a higher infall velocity and a broader profile. Indeed the radius of the HD1 core is only 91\% of the radius of the HD0 at the respective best-fit times.

\begin{figure}
\centering
        \includegraphics[width=1.0\columnwidth]{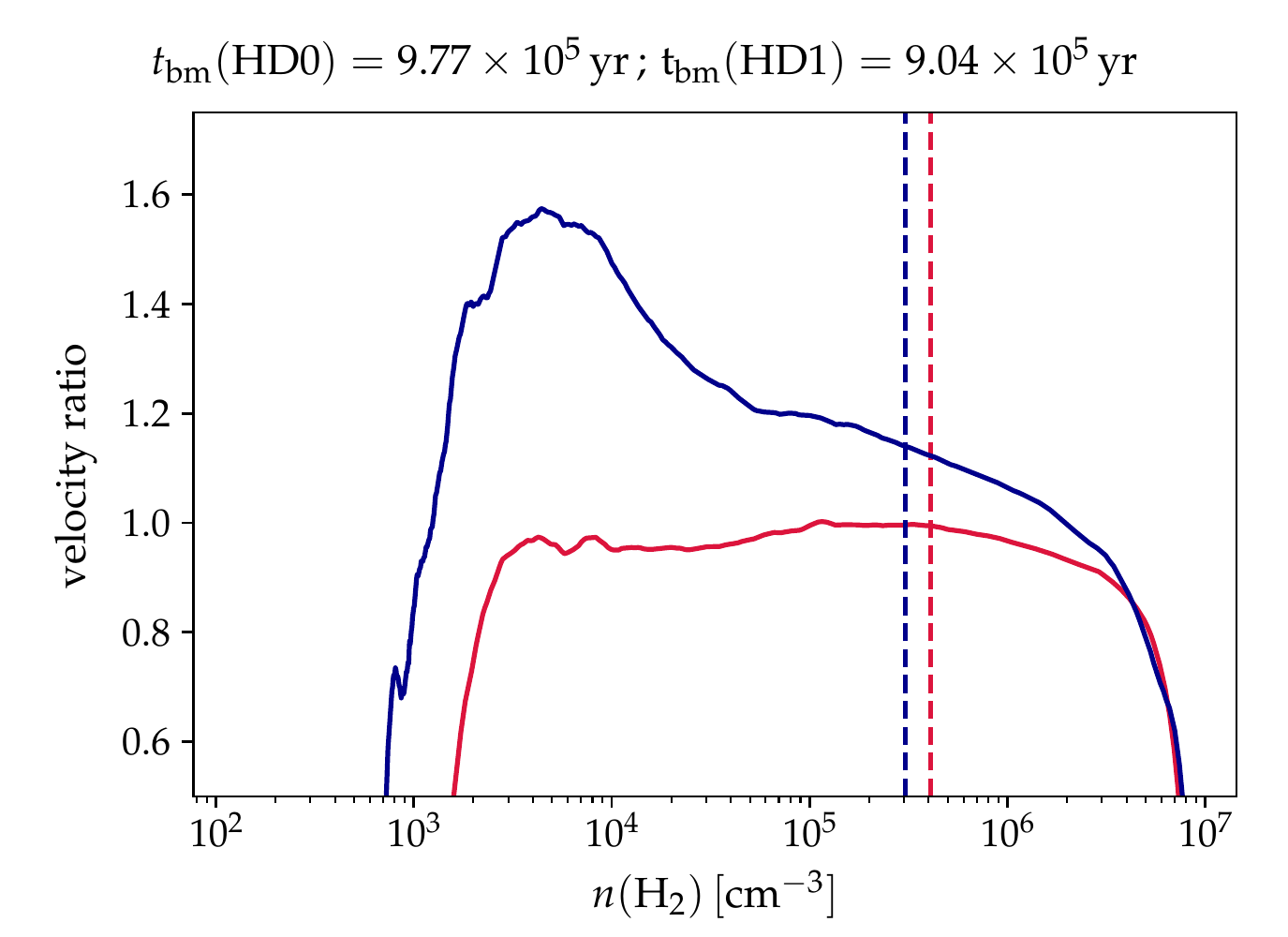}
    \caption{Ratios of the HD0 (red) and HD1 (blue) velocity profiles to the KC velocity profile as a function of volume density at the best-fit times in the HD0 and HD1 simulations, indicated above the figure. The positions of the infall velocity peak in the HD0 and HD1 simulations are shown as dashed vertical lines as a guide to the eye.}
    \label{fig:velocityRatio}
\end{figure}

Figure~\ref{fig:velocityRatio} shows the ratio of the velocity profiles in simulations HD0 and HD1, and in the KC model. Interestingly, the velocity profile in simulation HD0 is extremely similar to that in KC in the infalling regions. In simulation HD1, the velocity is however higher by a factor between 1 and 1.6 depending on the location in the core.

\subsection{Hydrodynamical simulations HD2 and HD3}\label{ss:alternativeModels}

\begin{figure*}
\centering
        \includegraphics[width=2.0\columnwidth]{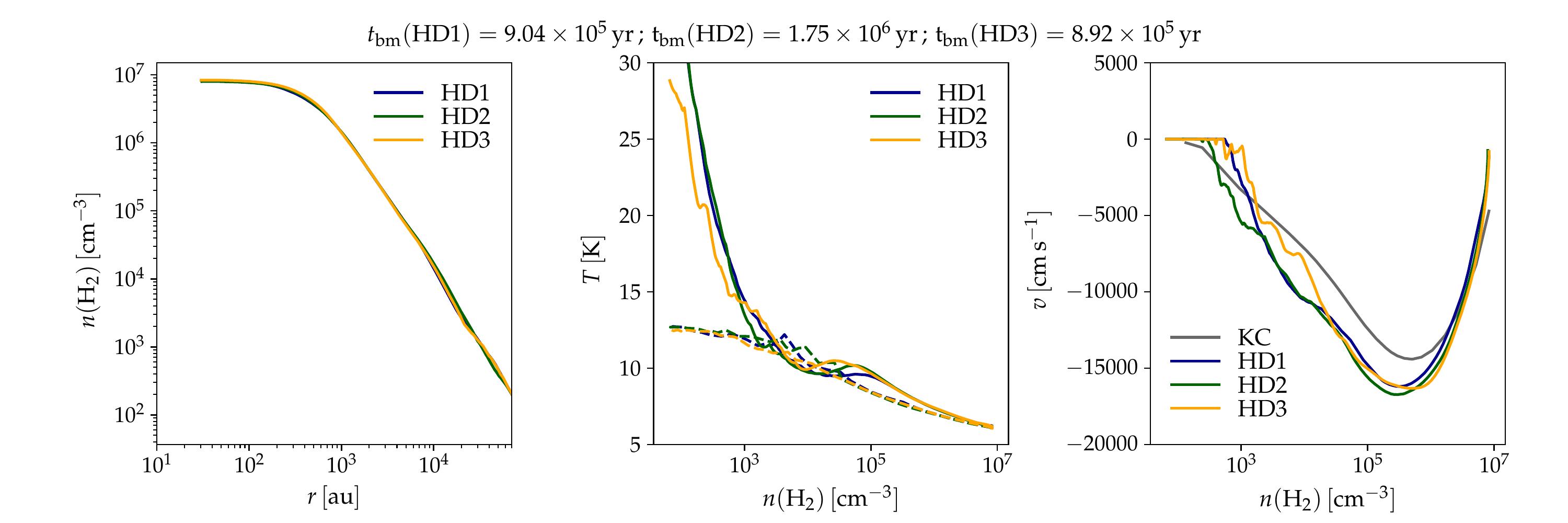}
    \caption{As Fig.\,\ref{fig:physicalStructureFiducial}, but comparing the results of simulations HD2 and HD3 to HD1. The KC infall velocity profile is shown in the right-hand panel as a reference.}
    \label{fig:KC_vs_H1}
\end{figure*}

We assess here the effects of complex chemistry and approach to line cooling on the simulation results. This is achieved by making further modifications to simulation~HD1, first simplifying the chemistry to the B2alt chemical network of S18 in simulation~HD2, and then switching to the parametrized line cooling scheme in simulation~HD3. The KC model employs the parametrized cooling powers of neutral molecular species from \citet{Goldsmith01}, as well as cooling by $\rm C^+$ and atomic O following \citet{Tielens05}\footnote{Atomic O is also included in the set of coolants considered by \citet{Goldsmith01}.}. The latter is based on treating the excitation and de-excitation as a two-level system \citep{Keto14}. We have in our simulation~HD3 (and in HD4) adopted the same parametrized treatment for cooling.

The results of simulations HD2 ($t_{\rm bm}({\rm HD2}) = 1.75 \times 10^6$\,yr) and HD3 ($t_{\rm bm}({\rm HD3}) = 8.92 \times 10^5$\,yr) are shown in Fig.\,\ref{fig:KC_vs_H1}. We also show the profiles from simulation~HD1 for reference, as in that one the core is not allowed to expand either. Evidently, there is no substantial difference between the results of the three simulations as far as the infall velocity is concerned, and all curves are very similar to each other. There are small differences in the temperature profiles, which is expected given the varying treatments to chemistry and cooling. All gas temperature and infall velocity profiles present a certain degree of fluctuation at low volume densities, which is an interpolation artifact -- the chemical and radiative simulations are resolved on a mesh of 35 points, and the results are interpolated onto the full hydrodynamical mesh consisting of 1000 points. A curious detail is that the gas temperature profile in simulation~HD3 does not resemble that of KC at low volume densities, and we do not recover a strong local decrease of the temperature around $n({\rm H_2}) = 10^3 \, \rm cm^{-3}$.

Finally we note that in all simulations that adopted the small chemical network, we had to enforce the no-expansion condition in order to obtain a collapsing solution. Otherwise the core starts to expand before the main coolants form and collapse does not occur. This does not mean that we expect collapse to never occur if one uses a simple chemical network. If we altered the parameters of the core model such that the photoelectric heating was less efficient in the outer core, and hence the expansion motion was weaker, it could be possible to obtain a collapsing solution without ad hoc modifications to the dynamics. We reiterate that the present choice of model parameters is motivated purely by the desire to obtain similar physical conditions to the KC model toward the end of the collapse.

\subsection{Hydrodynamical simulation HD4}

The simulation HD4, where the larger initial core is adopted, led to interesting results in that the core did not collapse within the maximum simulation duration of $10^7\,\rm yr$. During its evolution, the core experienced a few brief periods of weak collapsing motions which ended quickly, such that the central density increased only by a factor of 2.1 over $10^7\,\rm yr$. Table~\ref{tab:bestFitTimes} collects the best-match times in all of the five hydrodynamical simulations, as well as the corresponding free-fall times, calculated using the familiar relation
\begin{equation}
t_{ff} = \sqrt{\frac{3\pi}{32G\rho_{\rm ave}}} \, ,
\end{equation}
where $G$ is the gravitational constant and $\rho_{\rm ave}$ is the average density of the core. Evidently, simulations HD0 to HD3 all predict collapse in a time shorter than the free-fall time. The fact that the same does not occur in HD4 may indicate that even though the initial core is unstable (and isothermal), the chemical evolution and the associated introduction of a temperature profile in the early stages of the simulation cause the core to move toward the stable regime. However, applying the formulas for nonisothermal Bonnor-Ebert spheres presented in \citet{Sipila11}, we estimate that the nondimensional radius is only reduced in the early stages of the simulation from 50 to $\sim$33, which should still be well in the unstable regime -- though it needs to be noted that the analysis in \citet{Sipila11} extended only up to a core mass of 5\,$M_\sun$ under the assumption that the dust and gas temperatures are equal, which certainly does not hold in the present case. Therefore we cannot quantify without a dedicated analysis why the HD4 simulation develops differently, in terms of the collapse duration vs. free fall time, to the other four simulations. Also, we note that the age of the KC model core is $\sim$$6 \times 10^5 \, \rm yr$ when the central density hits $\sim$$10^7 \, \rm cm^{-3}$ \citep[see Appendix~D in][]{Caselli22}, while in our HD4 simulation, which is similar in setup to KC, the core does not collapse, at least within $10^7\,\rm yr$. We cannot offer a quantitative explanation for this difference without a direct comparison of the codes.

\begin{table}
        \renewcommand{\arraystretch}{1.3}
        \centering
        \caption{Best-match times $t_{\rm bm}$, defined by the closest match to the central density in the KC model, in the five hydrodynamical simulations presented in this paper. Also shown are the corresponding free-fall times.}
        \begin{tabular}{c | c | c}
                \hline
                \hline
                Simulation & $t_{\rm bm}\,\left[\rm yr\right]$ & $t_{\rm ff}\,\left[\rm yr\right]$ \\
                \hline
                HD0 & $9.77 \times 10^5$ & $2.70 \times 10^6$\\ 
                \hline
                HD1 & $9.04 \times 10^5$ & $2.70 \times 10^6$ \\
                \hline
                HD2 & $1.75 \times 10^6$ & $2.70 \times 10^6$ \\
                \hline
                HD3 & $8.92 \times 10^5$ & $2.70 \times 10^6$ \\
                \hline
                HD4 & -- & $6.38 \times 10^6$ \\
                \hline
        \end{tabular}
        \label{tab:bestFitTimes}
\end{table}

\subsection{Line simulations and comparison to observations}\label{ss:lineSimulations}

\begin{figure*}
\centering
        \includegraphics[width=2.0\columnwidth]{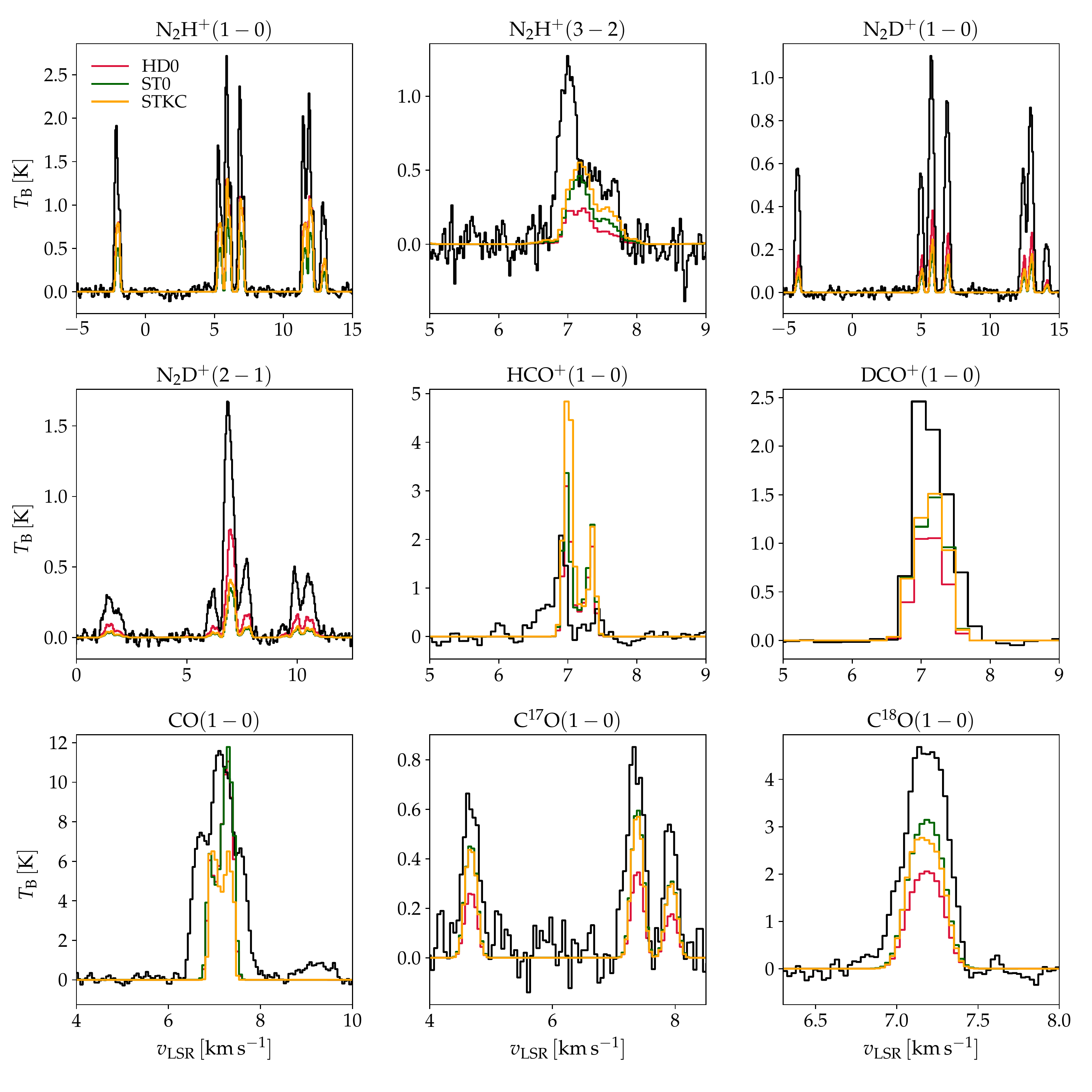}
    \caption{Simulated line profiles of various rotational transitions in simulations HD0 (red), ST0 (green), and STKC (orange). The black histograms show the observed lines.}
    \label{fig:lines}
\end{figure*}

\begin{figure*}
\centering
        \includegraphics[width=2.0\columnwidth]{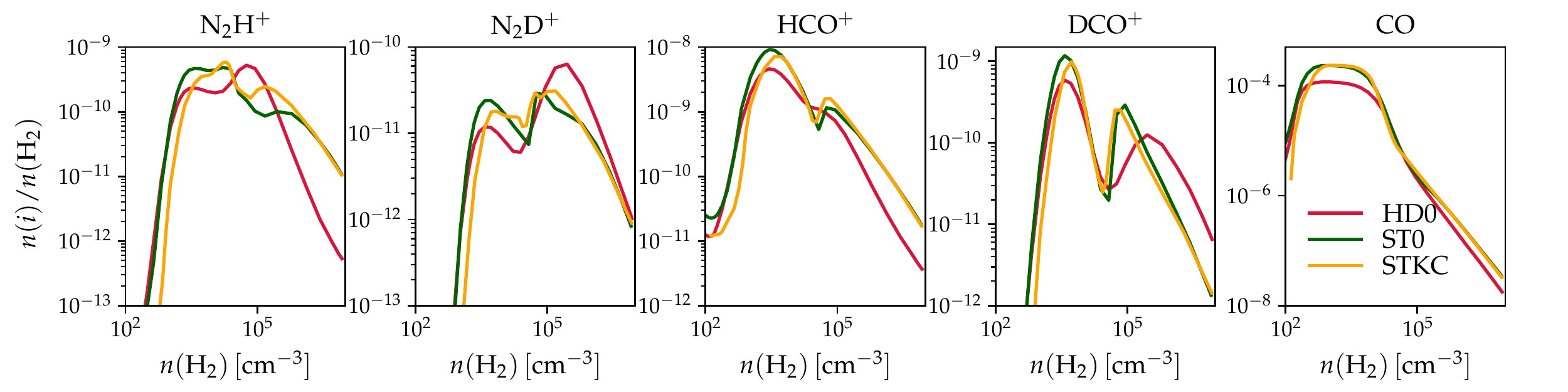}
    \caption{Abundances of the molecules whose line profiles are shown in Fig.\,\ref{fig:lines} as a function of volume density.}
    \label{fig:abundances}
\end{figure*}

We have run emission line simulations to constrain the effect of the infall velocity on observables. Figure~\ref{fig:lines} shows the so-obtained line profiles of several rotational transitions of $\rm N_2H^+$, $\rm N_2D^+$, $\rm HCO^+$, $\rm DCO^+$, CO, $\rm C^{17}O$, and $\rm C^{18}O$ in simulations~HD0, ST0, and STKC. The required input data for the line simulations, including the collisional rate coefficients, was obtained from the LAMDA database \citep{Schoier05}. For $\rm N_2H^+$ and $\rm N_2D^+$ we assume that the hyperfine components are distributed according to local thermodynamic equilibrium (LTE), with component separations and relative intensities calculated from the data of \citet{Pagani09b}. For $\rm DCO^+$, we use hyperfine-resolved rate coefficients from \citet{Pagani12}. All line simulations correspond to the time $t_{\rm bm}({\rm HD0})$. The figure also shows the  observed lines, previously presented in \citet{Chacon-Tanarro19a} and \citet{Redaelli19,Redaelli21}. To aid the interpretation of these results, we show in Fig.\,\ref{fig:abundances} the abundance profiles of the molecules as a function of volume density.

Comparing first the results of the HD0 and ST0 simulations reveals that the former produces more emission in $\rm N_2H^+(1-0)$, while the latter yields a brighter $(3-2)$ line. This can be understood in terms of the $\rm N_2H^+$ abundance. The $(1-0)$ line has a critical density of $2 \times 10^5 \, \rm cm^{-3}$ (at 10\,K), and around this density the $\rm N_2H^+$ abundance is higher in the HD0 model\footnote{We recall that the HD0 and ST0 simulations adopt the same gas temperature and volume density profiles, and so the excitation conditions are the same in the two cases.}. The critical density of the $(3-2)$, however, is $5 \times 10^6 \, \rm cm^{-3}$, and in this region of the core the ST0 model predicts a higher abundance and hence stronger emission. For $\rm N_2D^+$, the critical densities of the $(1-0)$ and $(2-1)$ are $1 \times 10^5 \, \rm cm^{-3}$ and $8 \times 10^5 \, \rm cm^{-3}$, respectively. It is thus natural that the $\rm N_2D^+$ lines are brighter in simulation HD0 where the abundance is boosted in the central core compared to simulation ST0. Both simulations clearly underpredict the observed emission in $\rm N_2H^+$ and $\rm N_2D^+$.

The peak intensity of the $\rm HCO^+ (1-0)$ (critical density $2 \times 10^5 \, \rm cm^{-3}$) is clearly overpredicted by both HD0 and ST0. The $\rm HCO^+$ abundance is slightly higher in the latter simulation, which translates to a higher peak brightness. The $\rm DCO^+(1-0)$ line (critical density $8 \times 10^4 \, \rm cm^{-3}$) originates in a region where the $\rm DCO^+$ abundance is higher in ST0, and hence the $(1-0)$ line is brighter in that simulation.

The CO abundance profile is very similar in the dynamical and static simulations, the abundance being slightly higher in the latter, and hence the $\rm ^{12}CO(1-0)$ line is very similar in the two simulations. The line is optically thick and traces the outer core only (critical density $2 \times 10^3 \, \rm cm^{-3}$). Our chemical model does not consider the isotope chemistry of oxygen, and therefore we have simply scaled the CO abundance profile to obtain an estimate of the $\rm C^{17}O$ and $\rm C^{18}O$ abundances. We used the following scaling factors, appropriate for the local interstellar medium: $\rm C^{17}O/CO$ = 1/1792, $\rm C^{18}O/CO$ = 1/560 \citep{Wilson94}. The difference in CO abundance between HD0 and ST0 becomes more evident in the $\rm C^{17}O$ and $\rm C^{18}O$ $(1-0)$ lines, and the latter simulation matches the observations better, though by no means well.

A comparison of the results of simulation STKC against HD0 (and ST0) is less straightforward because the former adopts the KC physical structure, and even though the density distribution is very similar to that in HD0, the spatial differences in the gas temperature and in the infall velocity lead to variations in the excitation and in the line shapes. In some instances, the STKC simulation predicts lines comparable to HD0, while in some other cases the results match better with ST0. In two cases, the STKC simulation predicts distinctly ``unique'' results: the $\rm ^{12}CO(1-0)$ line is strongly self-absorbed which is not seen in the HDCRT simulations, and the $\rm HCO^+(1-0)$ is clearly brighter compared to HD0 and ST0. The case of CO is interesting in that the observed profile, which is rather clearly caused by multiple velocity components along the line of sight, is matched somewhat better (but still badly) by the HD0 and ST0 simulations compared to STKC, despite the fact that the CO abundance is very similar between the three (Fig.\,\ref{fig:abundances}). This is suggesting that the infall velocity in the low-density outer core is overestimated in the KC model (see Sect.\,\ref{ss:abusAndInfall} for additional discussion regarding the outer core).

In terms of abundances, the overall differences between the dynamical and static models are in line with the predictions of S18, where it was found that the $\rm N_2H^+$ abundance is higher in static models than in dynamical models at volume densities $\sim$$10^4 \, \rm cm^{-3}$ and that the trend is reversed at higher densities. The present results show another reversal accompanied by very strong $\rm N_2H^+$ depletion for high volume densities in the dynamical model, which is seemingly in contradiction with S18 -- we note however that in that paper the central density of the model core only went up to $\sim$$5 \times 10^5 \, \rm cm^{-3}$, so a similar extremely efficient $\rm N_2H^+$ depletion is not seen in those results. Akin to S18, we obtain in the present work a higher degree of $\rm HCO^+$ depletion as well as stronger overall deuterium fractionation toward the center of the core in the dynamical model.

Table~\ref{tab:criticalDensities} collects the critical density of all simulated lines for reference. The Table also shows the location of the excitation temperature ($T_{\rm ex}$) peak of all the simulated lines. The locations of the $T_{\rm ex}$ peaks match quite well with the critical densities, showing that the critical density is a reasonably good indicator of the region where the line is most excited -- though the CO(1-0) line serves as a counterexample.

\begin{table}
        \renewcommand{\arraystretch}{1.3}
        \centering
        \caption{Characteristics of the simulated molecular lines.}
        \begin{tabular}{c | c | c }
                \hline
                \hline
                Transition & Critical density$^{(a)}$ & Location of $T_{\rm ex}$ \\
                 & $[ \rm cm^{-3}]$ & peak$^{(b)}$\,$[ \rm cm^{-3}]$ \\
                \hline
                $\rm N_2H^+(1-0)$ & $2 \times 10^5$ & $6 \times 10^5$\\
                $\rm N_2H^+(3-2)$ & $5 \times 10^6$ & $8 \times 10^6$\\
                $\rm N_2D^+(1-0)$ & $1 \times 10^5$ & $6 \times 10^5$\\
                $\rm N_2D^+(2-1)$ & $8 \times 10^5$ & $3 \times 10^6$\\
                $\rm HCO^+(1-0)$ & $2 \times 10^5$ & $2 \times 10^5$\\
                $\rm DCO^+(1-0)$ & $8 \times 10^4$ & $1 \times 10^5$\\
                $\rm CO(1-0)$ & $2 \times 10^3$ & $6 \times 10^2$\\
                $\rm C^{17}O(1-0)$ & $2 \times 10^3$ & $4 \times 10^3$\\
                $\rm C^{18}O(1-0)$ & $2 \times 10^3$ & $3 \times 10^3$\\
                \hline
        \end{tabular}
        \tablefoot{$^{(a)}$ Critical density of the transition as defined by the ratio of the Einstein coefficient and the collisional rate coefficient at 10\,K, without optical depth corrections. The quoted values are calculated based on the data files in the LAMDA database. $^{(b)}$ Location of the $T_{\rm ex}$ peak of the transition in terms of volume density at $t_{\rm bm}({\rm HD0})$.}
        \label{tab:criticalDensities}
\end{table}

\section{Discussion}\label{s:discussion}

It was found above that the abundances and infall velocity profile derived from the hydrodynamical models and the static simulation ST0 are quite similar to those predicted by the simulation adopting the KC physical structure (STKC). None of the models is able to reproduce the observed lines without modifications to the parameter space. We discuss these findings and their implications, as well as the uncertainties of the modeling setup, in what follows.

\subsection{Abundances and infall velocity}\label{ss:abusAndInfall}

It has been shown in \citet{Redaelli19} that the chemical abundance profiles obtained using the KC model -- when one adopts a set of ``standard'' parameters in the chemical simulation -- do not lead to a satisfactory match with the observations unless the abundance profiles are scaled by some factor. They considered different multiplicative factors that were used to scale the abundance of a given molecule across the core, and were in this way able to match approximately the observed line intensities, but not always the intensity ratios between several rotational transitions of the same molecule. The results presented in Fig.\,\ref{fig:lines} show that also in the present case, neither the dynamical model or the static models present a good overall match to the observations, and would require variations in the abundances (or in the physical model parameters) to reach a better agreement with the observed lines. Indeed, the predictions of chemical models are strongly dependent on a wide variety of parameters whose values are poorly known, and in addition there are sometimes large uncertainties in the rate coefficients of the most important reactions. There is however one distinct prediction where the dynamical model distinguishes itself from the static models: the prediction of a higher degree of depletion in nondeuterated species and higher degree of deuteration at the core center. It is possible that this effect could be used as a proxy to test whether the dynamical model is predicting the abundances better than the static models are, if (high-resolution) observations of nondeuterated and deuterated high-density tracers were compared to the results of the two types of model. An in-depth investigation of this, that is, to constrain the depletion factors and degree of deuteration in multiple molecules, requires however a multiline and multiscale observational campaign.

Concerning the inner core, \citet{Bizzocchi13} concluded that the infall velocity in the KC model must be scaled up by a factor of 1.75 to obtain the best match to $\rm N_2H^+(1-0)$ observed toward the dust peak of L1544, based on a $\chi^2$ minimization applied to simulations adopting either constant abundances or abundance gradients. \citet{FerrerAsensio22} on the other hand find via an analysis of the 1-0 line of $\rm HC^{17}O^+$ (critical density $2 \times 10^5 \rm cm^{-3}$, that is, very similar to that of $\rm N_2H^+(1-0)$) that a correction factor of 1.3 results in a good match with the observations. Our present model predicts very similar infall velocities to KC -- for the same central density -- which may suggest that the higher velocities implied by the line observations are the result of the central density of the core being higher than what is currently thought (see also Sect.\,\ref{ss:uncertainties}). Recent dust emission observations estimating the central density have led to contradictory results: the work of \citet{Chacon-Tanarro19b}, based on single-dish data, argues for a lower central density compared to KC, while the interferometric observations of \citet{Caselli19b} imply a central density of $\sim$$10^7 \, \rm cm^{-3}$. 

The positive gas velocity in the outer core in simulation~HD0 is due to photoelectric heating which causes the gas to expand. We assume in the hydrodynamical simulations that the properties of the gas just outside the core (which is not treated explicitly in the simulation) are the same as those at the edge of the model; there is no thermal pressure gradient across the boundary, so that such an expansion is a natural feature of the model when heating overcomes cooling. This result appears to be in contradiction with recent observational evidence on gas motions in the outer regions of L1544: $\rm HCO^+$ data presented by \citet{Redaelli22} indicate that the large-scale structure of L1544 is collapsing. However, it could be that the apparent discrepancy is due to radiative transfer effects, as lines of different molecules trace different layers of the gas. It is very difficult to explain the observations of gas motions at large scales with a model such as the present one, because in reality the gas velocities may vary with the direction outward from the core center, and a 1D spherically symmetric desription of the core is not appropriate for modeling such asymmetries.

\subsection{Uncertainties in the model parameters}\label{ss:uncertainties}

The simulations presented here suffer from uncertainties in a multitude of sources. Ignoring the obvious simplication arising from the use of a one-dimensional spherically symmetric model to represent a nonsymmetric object, error is introduced by uncertainties in the mass of the core, the central density, the gas and dust temperatures, and so on. We discuss in the following the various uncertainty sources and their impact on the conclusions of our work on a mostly qualitative level.

\begin{figure*}
\centering
        \includegraphics[width=2.0\columnwidth]{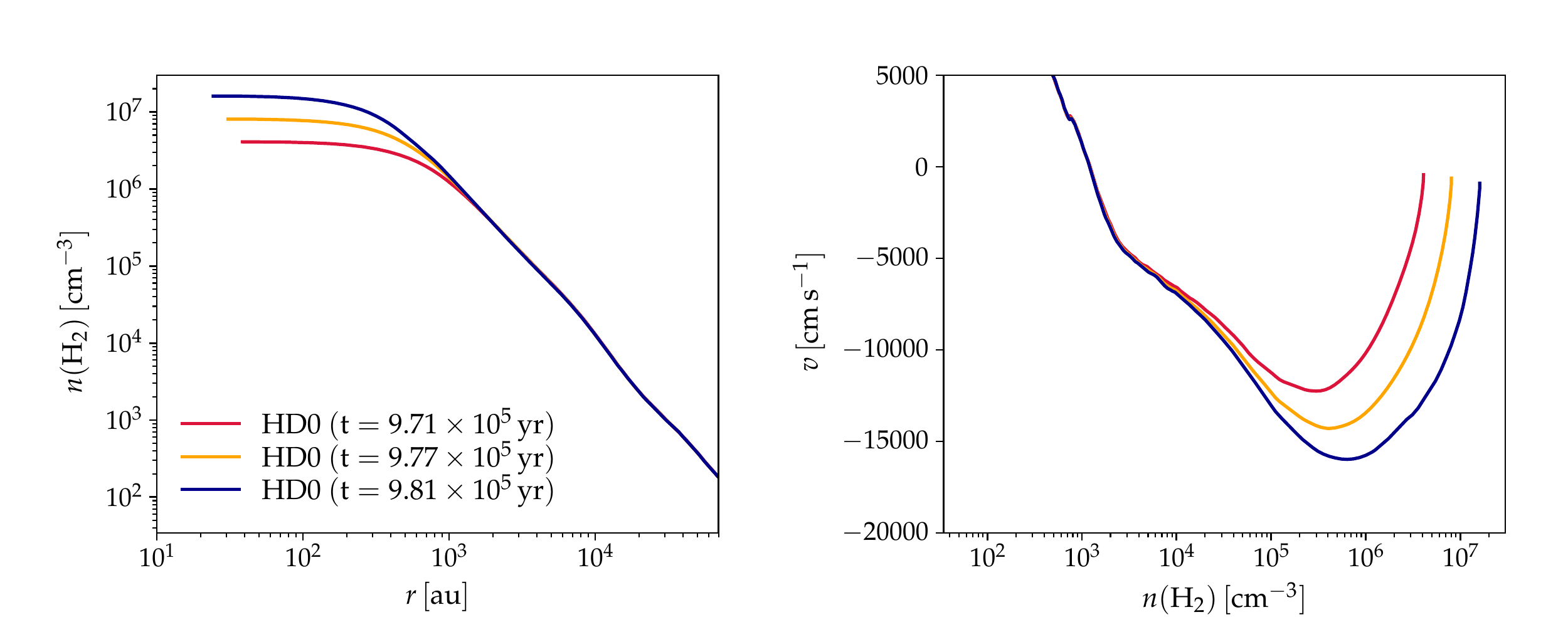}
    \caption{Physical properties of the HD0 core. {\sl Left:} Density profiles from the HD0 simulation at three different time steps, indicated in the Figure, as a function of radius. {\sl Right:} Infall velocity as a function of density at the three time steps displayed in the left panel.}
    \label{fig:physicalStructureDiffTimes}
\end{figure*}

We have set as the target central density, which determines the time step when the results of the hydrodynamical simulation are extracted, to correspond closely to the central density in the KC model ($8.3 \times 10^6 \, \rm cm^{-3}$). This central density estimate is based on a comparison with observations under the assumption that L1544 is at a distance of 140\,pc. The distance has since been updated to 170\,pc (\citealt{Galli19}; we adopt this value in the present paper), which means in practice that previous observations cover a larger area of the core than previously thought. The longer distance allows for a higher central density while still reaching the same average density as in the previous works, for a given beam area. A practical consequence of a higher (lower) central density for our simulation results would be a higher (lower) peak infall velocity, as well as an inward (outward) shift in the location of the peak. We show in Fig.\,\ref{fig:physicalStructureDiffTimes} the density and infall velocity profiles in the HD0 simulation at time steps when the central density is a factor of two lower or higher than the KC central density. Clear differences in the infall velocity profile are apparent at the three different times in the central areas of the core, and the peak infall velocity increases by about 23 per cent ($1.23 \times 10^4$ to $1.60 \times 10^4 \, \rm cm \, s^{-1}$ in absolute value) in just $10^4\,\rm yr$. The associated increase in $\rm H_2$ column density, calculated toward the center of the core assuming a 1000\,au (5.88$\arcsec$) beam, is only 44 per cent (increasing from $9.08 \times 10^{22}$ to $1.31 \times 10^{23} \, \rm cm^{-2}$) despite the factor of four change in central density. Constraining the central density via, for example, observations of dust emission with limited resolution is consequently very difficult.

\begin{figure*}
\centering
        \includegraphics[width=1.8\columnwidth]{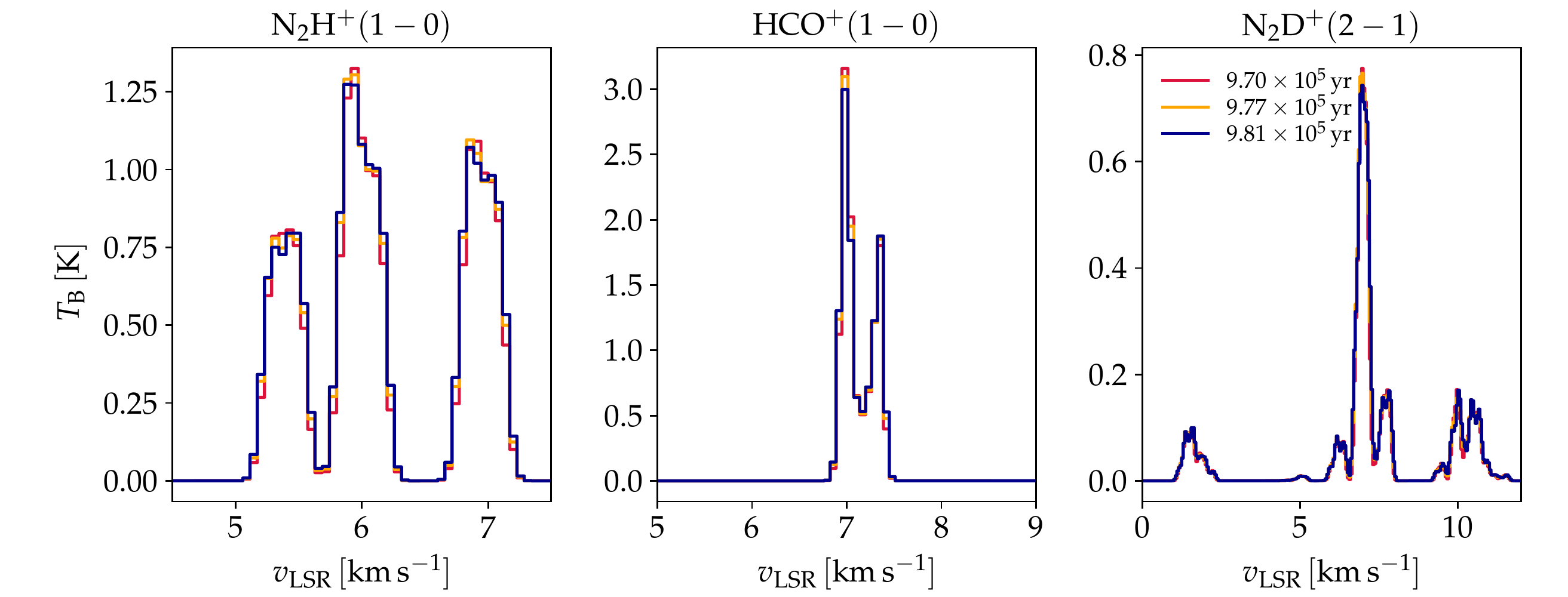}
    \caption{Line profiles of $\rm N_2H^+(1-0)$ main hyperfine group (left), $\rm HCO^+(1-0)$ (middle), and $\rm N_2D^+(1-0)$ (right) in simulation~HD0 computed at the three time steps depicted in Fig.\,\ref{fig:physicalStructureDiffTimes}. The lines are shown here without the comparison to the observed counterparts for clarity.}
    \label{fig:multipleTimes}
\end{figure*}

One is tempted to think based on Fig.\,\ref{fig:physicalStructureDiffTimes} that the line profiles of high-density tracers could be used to provide hints of the magnitude of the infall velocity, and hence of the central volume density. However, in practice this too may be difficult. We show in Fig.\,\ref{fig:multipleTimes} the line profiles of the two molecular transitions in our line selection which trace the central core and show definite signs of infall asymmetry, that is, $\rm N_2H^+(1-0)$ and $\rm HCO^+(1-0)$, in simulation~HD0 at the three different time steps. Neither line shows clear signs of changes as a function of time, although the absorption dip of the $\rm N_2H^+$ hyperfine component at 5.4\,km\,s$^{-1}$ does get very slightly deeper with time. The plot also shows the line profile of the $\rm N_2D^+(2-1)$ transition, which traces the gas near the center. Also in this case, we only see the effect of depletion in that the line intensity decreases slightly with time -- due to the low optical thickness of the line, the increasing infall velocity does not lead to prominent effects on the line shape.

As noted in Sect.\,\ref{ss:abusAndInfall}, the velocity profile in the KC model has been up-scaled globally in some previous works to reconcile simulation results with observations. We showed above that, in advanced stages of collapse, such global changes are not expected based on variations in the central density alone -- the associated infall velocity changes are limited to the inner core. Hence it may be that achieving a higher infall velocity on larger scales requires an alteration of the average density of the core. It is plausible that a higher average density would lead to a higher infall velocity, and this notion is supported by the comparison between our models H1 and H4. More precise estimations of the core mass, taking into account the updated source distance, are needed to provide constraints for future simulations. It is worth noting though that there are additional considerations which we cannot probe with the 1D simulation setup, such as ambipolar diffusion and flattening of the envelope, that are likely to affect the collapse dynamics. For discussion on these effects in the context of the evolution of prestellar cores, we refer the reader to \citet{Tassis12a,Priestley19,Tritsis22} for the former, and to \citet{Tritsis16} for the latter.

We have set the parameters of the radiation field incident on the core, controlled by the $G_0$ factor as well as the extinction external to the core, such that we recover at the end of the simulation a very similar central dust temperature to that in the KC model. In order to facilitate this, we carried out a series of test simulations using the KC physical structure before running the hydrodynamical simulations, so that we could find a set of parameters that will ultimately lead to the ``expected" dust temperature at the center of the collapsing core. This was achieved with external $A_{\rm V} = 1\, \rm mag$ and $G_0 = 0.5$. We emphasize that these parameter values were chosen for two reasons: 1) To obtain a good correspondence of our model with KC in terms of the dust temperature in the inner core, which requires $G_0 < 1$ with the dust model we employ here; and 2) To avoid excess heating of the gas in the outer core, which leads to efficient expansion and ultimately prevents the collapse of the inner core by transferring mass away from it. One further consideration here is the strength of the photoelectric heating effect. If $G_0$ and the external $A_{\rm V}$ were varied, one could still reach the target dust temperature by changing the dust opacity model. However, the dust model cannot be altered completely freely -- one must still obtain reasonable values for the photoelectric heating rates in the outer core. If the heating rate is too low, line cooling will completely dominate the thermal balance and the gas temperature drops to very low values. The practical problem in this regard is that the core model encompasses a large range of volume densities, and the optical properties of the dust cannot be expected to be the same on the one hand in the central core where the grains are coated with a thick ice, and on the other hand in the outer core where there is hardly any ice on the grains. This could be remedied by adopting a series of dust models where the optical properties change with depth into the core, but such simulations, while technically feasible, are out of the scope of the present work.

\section{Conclusions}\label{s:conclusions}

We presented the results of hydrodynamical simulations describing the gravitational collapse of a starless core whose properties (mass, outer radius) resemble those of L1544. Our aim was to investigate whether the adoption of a large chemical network and a time-dependent determination of the line cooling powers carried out via radiative transfer simulations could have an effect on the infall velocity of the gas -- specifically to increase it -- as compared to previous similar simulations where a very limited chemical network and a different approach to line cooling was considered \citep{Keto15}. We also ran so-called static simulations, where the physical structure of the core is kept fixed as the chemistry is evolving. We performed line simulations using the time-dependent chemical abundance profiles obtained from the hydrodynamical and static models, and compared the results to single-dish observations of various lines toward L1544.

We found that the present model predicts a very similar infall velocity compared to the earlier model for L1544 by \citeauthor{Keto15}\,(\citeyear{Keto15}; KC), who used a parametrized line cooling scheme. However, the time it takes for collapse to occur depends on the simulation setup. Using a large chemical network decreases the simulation duration by a factor of $\sim$ two, compared to simulations adopting a simple chemical network when the line cooling power is determined via radiative transfer means, in accordance with our previous work \citep{Sipila18}. This has obvious implications to chemical evolution, as demonstrated in \citet{Sipila18}, although that is not a point of focus in the present work. However, if we use a simple chemical network in connection with the parametrized line cooling scheme \citep{Goldsmith01,Tielens05}, the simulation duration is again reduced compared to a simulation using a simple chemical network but an on-the-fly radiative transfer treatment of line cooling. Determining the line cooling power self-consistently is clearly preferred over parametric expressions when estimating the lifetime of a core.

To facilitate the comparison of our results to the KC model, we extracted the abundances in our simulations at the time when the central density matches the one in KC, $n({\rm H_2}) = 8.3 \times 10^6 \, \rm cm^{-3}$. As the present simulations cannot explain the double-peaked nature of some transitions observed toward L1544, which requires higher infall velocities than previously suggested by simulations, we suggest that either the central density of L1544 is higher than it has been thought before -- in other words, the core may be dynamically more evolved -- or the estimation of the average density of the core is off. We investigated the first option, and found that an increase of a factor of two in the central density, which takes a few kyr when the core is in an advanced state of collapse, corresponds to an increase in infall velocity in the ten per cent level only, and that the increase in the infall velocity is limited to the inner core. The molecular lines in our sample that trace the very inner core are very insensitive to such changes in the infall velocity in the inner few thousand au. Observations have implied that the infall velocity is expected to be a factor of 1.3 to 1.75 higher than in the KC model, which has been deduced by scaling the infall velocity globally. Achieving a globally higher infall velocity would likely require increasing the average density of the simulated core, which we do not attempt in the present work.

Our results show that the hydrodynamical simulation, where the physical structure and chemical abundances both evolve with time, is in general better at reproducing the observed lines as compared to the static models where the core structure is fixed. Given that the simulation duration is very different (factor of $\sim$2) when the chemical description is simplified leads us to conclude that the effect of chemistry on the dynamics of the gravitational collapse of star-forming cores should not be downplayed -- one needs to consider the chemistry to the fullest possible extent in order to obtain a realistic picture of chemical evolution during the collapse, especially when paired with a self-consistent simulation of line cooling powers. Post-processing, that is, calculating the chemical abundances using the core structure from previous hydrodynamical simulations with simplified chemistry, is certain to yield different results compared to a fully self-consistent simulation if the difference in evolutionary timescale is on the order of 2. This holds even if the ``final'' physical state of the core is similar between the full and simplified simulations, as our present results show it to be.

The present simulations are far from perfect when it comes to predicting the observed line profiles toward L1544. Here we have intentionally refrained from trying to obtain the best possible fit to the observations, so that the effect of the changes in the infall velocity profile can be isolated and quantified better. Still, it is not reasonable to expect that a one-dimensional model such as the one presented here could provide a good match to all observed lines, especially for transitions arising in the outer core where the geometry of L1544 is clearly nonspherical -- though the presently used one-dimensional physical structure is strikingly similar to that originating in a three-dimensional model along the major axis of the core \citep{Caselli22}. However, further work can be done to understand better still the interaction of chemistry and the collapse dynamics, as there are several time-dependent effects not considered in the present model that impact the abundances of the various cooling molecules and hence the gas temperature, as well as the infall velocity. Two examples of these are the revised description of cosmic ray induced desorption presented by \citet{Sipila21}, which we did not consider here in order to remain maximally consistent with the work of \citet{Keto15}, and a time-dependent treatment of chemical desorption \citep{Vasyunin17,Riedel22}. Expanding the simulation setup will help to constrain the chemical evolution in the inner core in particular.

\begin{acknowledgements}
The authors acknowledge the financial support of the Max Planck Society.
\end{acknowledgements}

\bibliographystyle{aa}
\bibliography{AA43935.bib}

\begin{appendix}

\section{Optimization of the model}\label{a:optimization}

We are presently interested in modeling the gravitational collapse of a star-forming cloud using a geometrically simple 1D model framework. In the previous version of HDCRT, the time step used in the chemical and radiative transfer simulations was tied to the hydrodynamical time step, so that the dust temperature was also updated at every time step. However, once the simulation starts, the cloud spends a relatively long period of time in a quiescent phase where the density profile is almost unchanging. Because the dust opacity model or the strength of the radiation field external to the cloud do not change during the simulation, any temporal changes in the dust temperature depend mainly on changes in the gas density profile. Therefore there is in actuality little need to constantly update the dust temperature, and indeed one can save a significant amount of computational time by decreasing the frequency at which the dust temperature profile is determined. A similar argument applies to the chemical simulations; chemical evolution is slow in the outer areas of the core where the density is low, and a sparser time resolution is sufficient compared to the dense inner core where evolution is more rapid, while still obtaining a sufficiently accurate solution.

For the present work, we have introduced a multiple time step approach in HDCRT: the hydrodynamical, dust temperature, and chemical simulations are run at variable time intervals. The solution to the hydrodynamics proceeds on a time step $\Delta t_{\rm hydro}$, determined by the Courant condition (see S18), and this represents the lower limit for a time step of any type. The frequency of the dust temperature simulation $\Delta t_{\rm dust}$ is defined stepwise, and is tied to the central density of the core. In the initial, quiescent stages of core evolution, the dust temperature is updated every $5\times10^3 \, \rm yr$ (central density $2 \times 10^4 \, \rm cm^{-3}$). The interval is decreased with increasing central density. When the central density climbs above $10^5 \rm cm^{-3}$, $\Delta t_{\rm dust} =  \max(\Delta t_{\rm hydro},10^2\,\rm yr)$ -- as noted above, the simulation is designed to terminate if $\Delta t_{\rm hydro}$ ever falls below $10^2\,\rm yr$. Using the variable time step method, one can save computational time in the initial stages of the simulation while also making sure that the dust temperature is updated often enough as the core begins to collapse. We note that a similar optimization scheme has been previously discussed by \citet{Keto05}.

\begin{figure}
\centering
        \includegraphics[width=1.0\columnwidth]{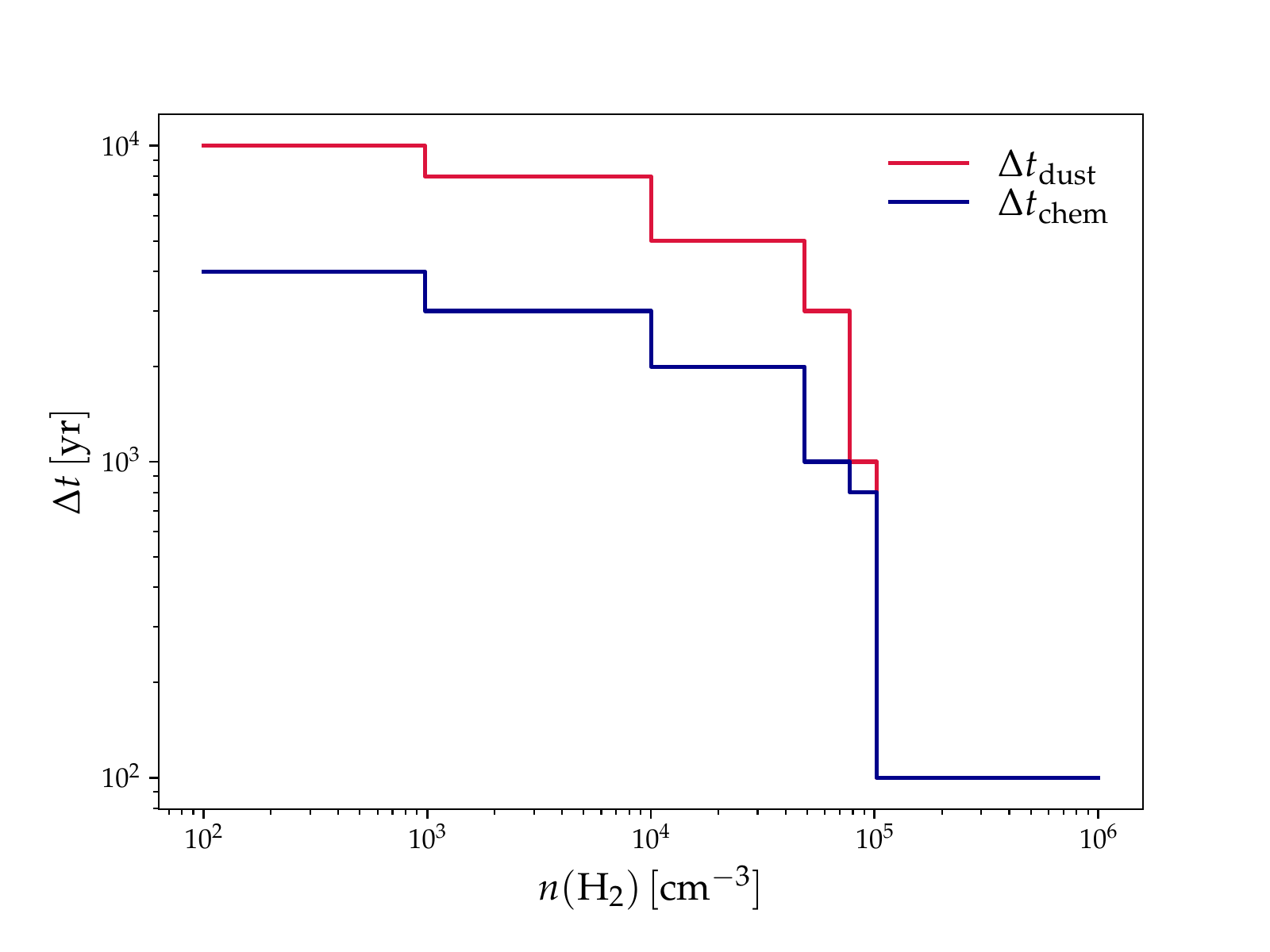}
    \caption{Minimum update intervals of the dust temperature ($\Delta t_{\rm dust}$) and chemistry ($\Delta t_{\rm chem}$) simulations as a function of $\rm H_2$ density.}
    \label{fig:timeSteps}
\end{figure}

The chemical simulation time step $\Delta t_{\rm chem}$ is defined differently to $\Delta t_{\rm hydro}$, due to the fact that the speed of the chemical evolution does not only depend on the central density of the core, but is rather a function of location\footnote{The temperature plays a part in the speed of chemical evolution as well.}. Chemical simulations are run in each model cell at the initial time step, but are subsequently updated in each cell only at intervals that depend on the volume density in a given cell. For example at early times in the simulation, the chemical simulations are updated every $2 \times 10^3 \, \rm yr$ in the cells near the center, while those near the edge are updated every $4 \times 10^3 \, \rm yr$. The number of cells in each density bin varies with time as the cells move. On the practical level, the code includes a series of counters that keep track of the necessity of updating the chemical simulations in each cell. The net effect of this arrangement is that, usually, chemical abundances are updated only in a subset of the cells at a given hydro time step, which saves a significant amount of computational time due to the heavy nature of the chemical simulations -- however at the cost of accuracy. We have determined through a series of tests with single-point chemical models that the values of $\Delta t_{\rm chem}$ shown in Fig.\,\ref{fig:timeSteps} produce accurate results when compared to simulations using smaller time steps.

The simulation of line cooling via radiative transfer (carried out using the LOC program; \citealt{Juvela20}) is a critical part of the HDCRT workflow: the line cooling powers of various molecules depend on the temporally and spatially varying abundances of the cooling molecules. The abundance profiles vary at every hydro time step, even in the case that the chemical simulations are run only in a subset of the cells at a given time step, and hence it is necessary to run the line cooling simulations at full temporal resolution, that is, following $\Delta t_{\rm hydro}$. Computational time savings on this part of the model framework can therefore be achieved only via optimizations to the radiative transfer simulations.

We note that the chemical simulation optimizations described above are possible thanks to the Lagrangian nature of the hydrodynamical model. In grid models, one must take into account advection between the model cells, and hence similar optimization schemes would likely be extremely detrimental to the accuracy of the simulation.

\end{appendix}

\end{document}